\documentclass{ws-rmta}
\usepackage{graphicx}
\usepackage[francais,english]{babel}
\usepackage[latin1]{inputenc}
\usepackage[T1]{fontenc}
\usepackage{amsmath}
\usepackage{amssymb}
\usepackage{color}
\usepackage{psfrag}
\usepackage{graphics}
\usepackage{mathrsfs}
\usepackage{verbatim}
\usepackage{textcomp}
\usepackage{xspace}
\usepackage{url}

\newcommand{\braket}[2]{\ensuremath{\langle #1 | #2\rangle}\xspace}

\begin{document}

\markboth{}{}

\catchline{}{}{}{}{}

\title{Finite $N$ corrections to the limiting distribution of the smallest eigenvalue of Wishart complex matrices}

\author{Anthony Perret}

\address{Laboratoire de Physique Th\'eorique et Mod\`eles  Statistiques, Universit\'e Paris-Sud, B\^at. 100, 91405 Orsay Cedex, France
\email{\footnote{anthony.perret@u-psud.fr}} }

\author{Gr\'egory Schehr}

\address{Laboratoire de Physique Th\'eorique et Mod\`eles  Statistiques, Universit\'e Paris-Sud, B\^at. 100, 91405 Orsay Cedex, France
\email{\footnote{gregory.schehr@u-psud.fr}} }

\maketitle

%\begin{history}
%\received{(Day Month Year)}
%\revised{(Day Month Year)}
%\end{history}

%\author{Anthony Perret \and Gr\'egory Schehr}
%\institute{A. Perret \at Laboratoire de Physique Th\'eorique et Mod\`eles  Statistiques, Universit\'e Paris-Sud, B\^at. 100, 91405 Orsay Cedex, France\\ %\and G. Schehr \at  Laboratoire de Physique Th\'eorique et Mod\`eles  Statistiques, Universit\'e Paris-Sud, B\^at. 100, 91405 Orsay Cedex, France}   
%\email{anthony.perret@u-psud.fr}
%\email{gregory.schehr@u-psud.fr}
 
%\date{\today}

\begin{abstract}
We study the probability distribution function (PDF) of the smallest eigenvalue of Laguerre-Wishart matrices $W = X^\dagger X$ where $X$ is a random $M \times N$ ($M \geq N$) matrix, with complex Gaussian independent entries. We compute this PDF in terms of semi-classical orthogonal polynomials, which are deformations of Laguerre polynomials. By analyzing these polynomials, and their associated recurrence relations, in the limit of large $N$, large $M$ with $M/N \to 1$ -- i.e. for quasi-square large matrices $X$ -- we show that this PDF, in the hard edge limit, can be expressed in terms of the solution of a Painlev\'e III equation, as found by Tracy and Widom, using Fredholm operators techniques. Furthermore, our method allows us to compute explicitly the first $1/N$ corrections to this limiting distribution at the hard edge. Our computations confirm a recent conjecture by Edelman, Guionnet and P\'ech\'e.  We also study the soft edge limit, when $M-N \sim {\cal O}(N)$, for which we conjecture the form of the first correction to the limiting distribution of the smallest eigenvalue.  
\end{abstract}

\section{Introduction}\label{section:Intro}

The study of extreme eigenvalue statistics in Random Matrix Theory (RMT) has attracted much attention during the last twenty years. In particular, the Tracy-Widom (TW) distributions \cite{TW94a,TW96} describing the largest eigenvalue $\lambda_{\max}$ (as well as the smallest one, $\lambda_{\min}$) in the classical Gaussian ensembles, orthogonal (GOE, $\beta =1$), unitary (GUE, $\beta= 2$) and symplectic (GSE, $\beta =4$) -- where $\beta$ is the Dyson index -- have become cornerstones of the theory of extreme value statistics of strongly correlated variables. Quite remarkably, it was shown that the TW distributions, denoted by ${\cal F}_{\beta}$ in the following, appear in a wide variety of problems \cite{Majumdar06}, a priori not directly related to RMT, ranging from the longest increasing sequence of random permutations of integers \cite{Baik99}, stochastic growth and related directed polymer models in the Kardar-Parisi-Zhang \cite{KPZ86} universality class \cite{Johansson00,Prahofer00,Sasamoto10,Calabrese10,Dotsenko10,Amir11} and sequence alignment problems \cite{Majumdar05} to non-intersecting interfaces and Brownian motions \cite{Nadal09,Forrester11,Liechty12} as well as in finance \cite{Biroli07}. While the TW distribution describes the typical fluctuations of $\lambda_{\max}$ and $\lambda_{\min}$, a large body of work has also been devoted to the study of large deviations of extreme eigenvalues in Gaussian ensembles \cite{Majumdar14}.

\begin{figure}[h!]
\begin{center}
\resizebox{120mm}{!}{\includegraphics{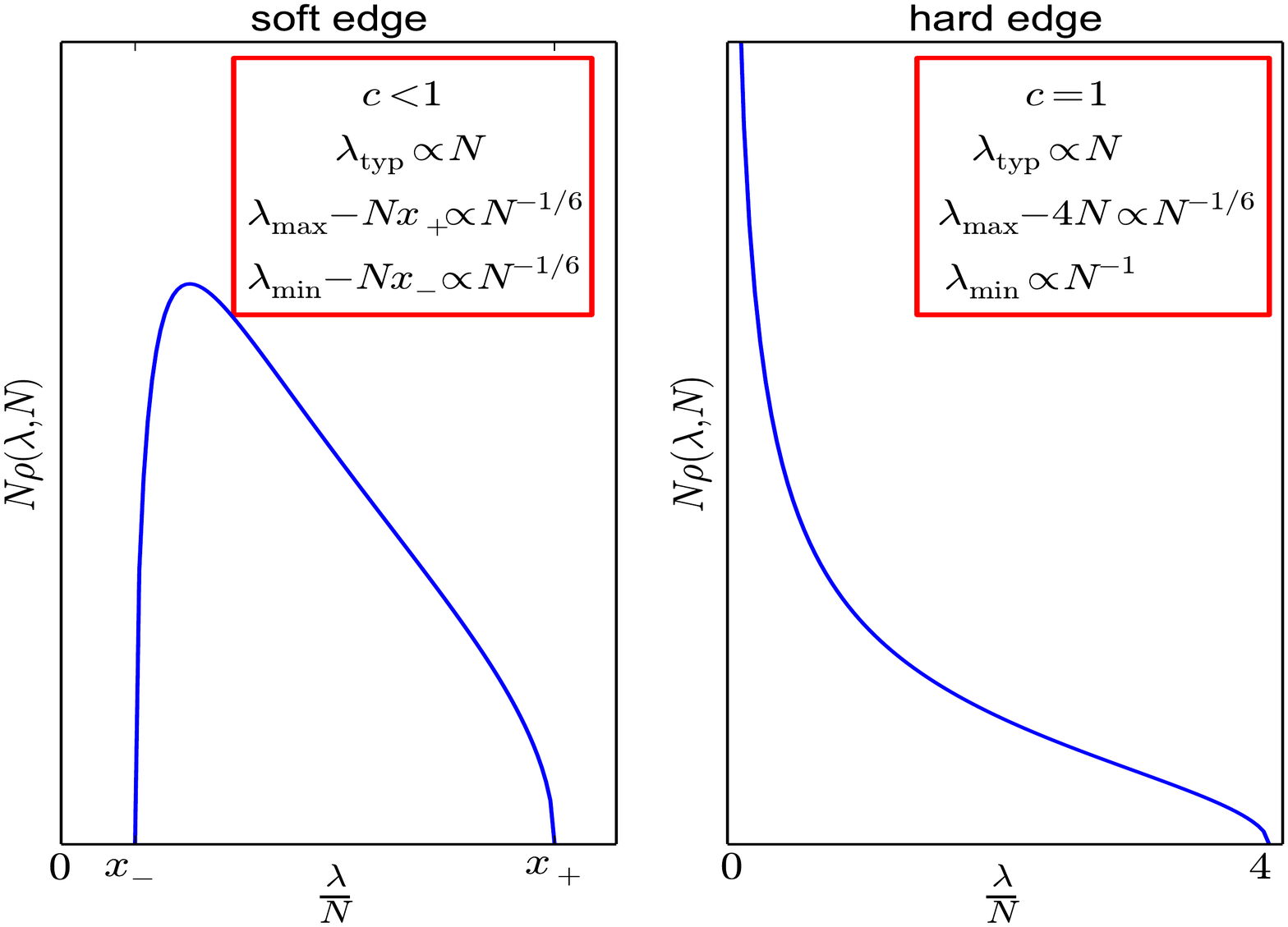}}
\caption{Plot of the Marchenko-Pastur distribution for $c<1$ ({\bf left}) and for $c=1$ ({\bf right}).}
\label{fig:Wishart_soft_vs_hard}
\end{center}
\end{figure}

Another interesting ensemble of random matrices, which we focus on in this paper, is the so-called Wishart-Laguerre ensemble -- here we focus on the case of complex matrices ($\beta = 2$). Let $X$ be a $M \times N$ rectangular matrix with i.i.d. complex Gaussian entries and $M-N=a  \geq 0$. The Wishart-Laguerre matrix $W$ is defined as $W=X\,^\dagger X$ which is thus a $N \times N$ Hermitian matrix, having $N$ real and positive eigenvalues $\lambda_1,\lambda_2,\cdots,\lambda_N$. The joint probability density function (PDF) of these $N$ eigenvalues is given by \cite{Mehta91,Forrester10}  
\begin{eqnarray}\label{Pjoint}
P_{\rm joint}(\lambda_1,\lambda_2,\cdots,\lambda_N)=\frac{1}{Z_N} \prod_{i < j}^N (\lambda_i-\lambda_j)^2 \,\exp{\left(-\sum_{i=1}^N \lambda_i\right)} \prod_{i=1}^N\lambda_i^a,
\end{eqnarray}
where $\lambda_1$, $\lambda_2$ ... $\lambda_N$ are positive and where $Z_N$ is a normalization constant, depending on $a$. These Wishart-Laguerre matrices play an important role in statistics, in particular in principal component analysis, where the matrix $W$ is a covariance matrix. Hence in this case both $M$ and $N$, and thus $a = M-N$, are positive integers. However, this joint PDF in Eq. (\ref{Pjoint}) is well defined for any real value of $a \geq 0$, which, for some non-integer value of $a$, may have some physical applications. An interesting example is the case of non-intersecting Brownian excursions \cite{Tracy07}, i.e., $N$ non-colliding positive Brownian paths $x_i(t) \ge 0$ on the unit time interval $t\in[0,1]$ constrained to start and end at the origin $x_i(0)=x_i(1)=0$. The joint PDF of the positions of the $N$ walkers, at a given time $t$, can indeed be written as \cite{Schehr08} 
\begin{eqnarray}\label{pdf_watermelons}
P_{\rm joint}(x_1,\cdots,x_N;t)=\frac{1}{z_N(t)} \prod_{i =1}^N x_i^{2} \prod_{i < j}^N (x_i^2-x_j^2)^2 \,\exp{\left(-\sum_{i=1}^N \frac{x_i^2}{\sigma^2(t)} \right)} \;,
\end{eqnarray}
where $\sigma^2(t) = 2t(1-t)$ and $z_N(t)$ is a normalization constant. From this expression (\ref{pdf_watermelons}) we obtain that the scaled variables 
$x_i^2/\sigma^2(t)$ behave statistically like the eigenvalues of random matrices from the Wishart-Laguerre ensemble (\ref{Pjoint}) with a non-integer parameter $a=1/2$. Hence it is physically relevant to study the joint distribution in Eq.~(\ref{Pjoint}) for any real $a \geq 0$.  

A first important characteristic associated to this ensemble (\ref{Pjoint}) is the mean density of eigenvalues, $\rho(\lambda,N)$, defined by
\begin{eqnarray}
\rho(\lambda,N)=\frac{1}{N} \sum_{i=1}^{N} \overline{\delta(\lambda_i-\lambda)},
\end{eqnarray}
where the overline denotes an average over the different realizations of the random variables $\lambda_i$'s according to the joint PDF in Eq.  (\ref{Pjoint}).
In the large $N$ limit, it is well known that $\rho(\lambda,N)$ is given by the Marchenko-Pastur (MP) distribution (see Fig. \ref{fig:Wishart_soft_vs_hard}) 
\begin{eqnarray}\label{density_MP}
\rho(\lambda,N) &\underset{N \to \infty}{\to}& \frac{1}{N}\rho_{\rm MP}\left( \frac{\lambda}{N}\right),\\
\rho_{\rm MP}\left(x\right)&=&\frac{\sqrt{(x-x_-)(x_+-x)}}{2 \pi x} ,
\end{eqnarray}
where $x_\pm = (c^{-1/2} \pm 1)^2$ are the right and left edges of the support, with $c=N/M \le 1$. The case where $c<1$ corresponds to the case where $a \sim {\cal O}(N)$. Here, we mainly focus on the case $c=1$ which corresponds instead to the case where $a$ is finite, while both $N$ and $M$ are large. In this case, which we will mainly focus on in this paper, the MP distribution takes the particular form (see the left panel of Fig.~\ref{fig:Wishart_soft_vs_hard})
\begin{eqnarray}\label{eq:MP_afinite}
\rho_{\rm MP}\left(x\right)&=&\frac{1}{2\pi}\sqrt{\frac{4-x}{x}} \;.
\end{eqnarray}
At the right edge, near $x_+ = 4$, $\rho_{\rm MP}\left(x\right)$ vanishes as a square-root, $\rho_{\rm MP}\left(x\right) \propto \sqrt{4 - x}$ and therefore the fluctuations near this {\it soft} edge are governed, for large $N$, by the Airy kernel. In particular, the distribution of the largest eigenvalue $\lambda_{\max}$, appropriately shifted and scaled, converges, when $N \to \infty$, to the TW distribution ${\cal F}_2$ mentioned above (the same as for GUE) \cite{Johansson00,Johnstone01}. It is now well known that this distribution can be expressed in terms of a special solution of a Painlev\'e II equation. While this connection to Painlev\'e transcendents was initially obtained by Tracy and Widom using Fredholm operator techniques, Nadal and Majumdar \cite{Nadal11} provided, more recently, a derivation of this result (for $\beta =2$) using semi-classical orthogonal polynomials (OPs), see also Ref. \cite{Chen05}. This method is at the heart of the present paper.  

On the other hand, at the left edge, $x_- = 0$, $\rho_{\rm MP}\left(x\right)$ has a square-root singularity, $\rho_{\rm MP}\left(x\right) \propto 1/\sqrt{x}$ (see the left panel of Fig. \ref{fig:Wishart_soft_vs_hard}). What about the fluctuations of the smallest eigenvalue $\lambda_{\min}$ in this case? One can estimate the typical scale of $\lambda_{\min}$, for large $N$, by considering that there is typically one eigenvalue in the interval $[0,\lambda_{\min}]$, i.e.,
\begin{eqnarray}
N \int_0^{\lambda_{\min}} \rho\left(\lambda,N\right) \, {\rm d} \lambda \sim {\cal O}(1)\;,
\end{eqnarray}
which implies that $\lambda_{\min} \sim {\cal O}(1/N)$. Within this scale ${\cal O}(1/N)$, the fluctuations are governed by the Bessel kernel \cite{For93,Ver93}. Furthermore, it has been shown that the distribution of $N \lambda_{\min}$ 
converges to a limiting form which (i) is different from the standard TW distribution ${\cal F}_2$ and depends continuously on the exponent $a$ in (\ref{Pjoint}) and (ii) can be written in terms of a special solution of a Painlev\'e III equation~\cite{Tracy94}. If one introduces $F_N(t)={\Pr}(\lambda_{\min} \ge t)$ then, one has indeed
\begin{eqnarray}\label{eq:result_TW}
\underset{N \to \infty}{\lim}F_N\!\left(\frac{x}{N}\right) = F_{\infty}(x),\hspace{0.2cm} F_{\infty}(x)=\exp\left( \int_0^x \frac{f(u)}{u} {\rm d}u\right) \;,
\end{eqnarray}
where $f(x)$ is the unique solution of a Painlev\'e III~\cite{Tracy94}:
\begin{eqnarray}\label{PIII}
(xf'')^2+4f'(1+f')(xf'-f)=(a f')^2 \;,
\end{eqnarray}
satisfying
\begin{eqnarray}\label{asympt_PIII}
f(x) \sim  - \frac{x^{a+1}}{\Gamma(a+1) \Gamma(a+2)}  \;, \;  {\rm as} \; x \to 0 \;.
\end{eqnarray}
This result (\ref{eq:result_TW}) was shown by Tracy and Widom using Fredholm operator techniques~\cite{Tracy94}. Note that for integer values of $a$, the limiting distribution $F_{\infty}(x)$ can be written as an $a \times a$ determinant whose entries are expressed in terms of Bessel functions (see Eq. (\ref{fdetBessel}) below) -- a result which can be obtained by clever manipulations of determinants~\cite{Forrester94}. In particular for $a=0$ the result is extremely simple as $F_N(x) = \exp{(-N\,x)}$ for all $N$, implying $f(x) = -x$, which is obviously solution of Eq. (\ref{PIII}) with the boundary condition~(\ref{asympt_PIII}).

The limiting distribution of $\lambda_{\min}$ for complex Wishart matrices, in the limit $N \to \infty$, is thus well known (\ref{eq:result_TW}) -- we refer the reader to Ref. \cite{akemann} to a recent work on the smallest eigenvalue for real Wishart matrices in the hard edge limit. What about the finite $N$ corrections to this asymptotic form? Such a question is quite natural for practical applications of extreme value statistics (EVS), where one always deals with finite samples -- here matrices of finite size. This issue was recently revisited for EVS of independent and identically distributed random variables using a renormalization group approach \cite{Gyorgyi10}. For EVS of strongly correlated variables, there are actually few cases where these corrections have been worked out explicitly, including random walks \cite{Schehr06}, the {\it largest} eigenvalue of random matrices belonging to various ensembles \cite{karoui2006,johnstone2008,johnstonema2011,Ma12}, non-intersecting Brownian motions \cite{Schehr13} or (Poissonized) random matchings \cite{Baik13}. For {\it real} Wishart matrices, the first corrections to the limiting distribution of the {\it smallest} eigenvalue in the soft edge limit were studied in Ref. \cite{Ma12} where it was shown that corrections to the limiting distribution of $\lambda_{\min}$ and $\lambda_{\max}$ are quite different, although the limiting distributions for both observables are actually the same, namely the TW distribution for GOE, ${\cal F}_1$. What about the corrections to the limiting distribution $F_{\infty}$ in Eq. (\ref{eq:result_TW}) of $\lambda_{\min}$ for complex Wishart matrices in the hard edge limit? This question was recently raised by Edelman, Guionnet et P\'ech\'e \cite{Edelman14} in their study of finite size covariance matrices with non-Gaussian entries. Based on the large $N$ expansion of the exact formulas obtained in Ref.~\cite{Forrester94} for small integer values of $a$, they conjectured the following form of the first $1/N$-correction
\begin{eqnarray}\label{conjecture_finite_N}
F_N\left(\frac{x}{N}\right) = F_\infty(x) + \frac{a}{2N} x \, F'_\infty(x) +  o\!\left(\frac{1}{N}\right) \;.
\end{eqnarray}
Note that this first $1/N$ correction in Eq. (\ref{conjecture_finite_N}) can be interpreted as a correction to the width, i.e., 
\begin{eqnarray}\label{eq:correction}
F_N\left(\frac{x}{N}\right) = F_{\infty}\left(x\left(1 + \frac{a}{2\,N}\right)\right) +  o\!\left(\frac{1}{N}\right) \;.
\end{eqnarray}
It is interesting to notice \cite{Baik13} that for most of the cases which have been studied in RMT~\cite{karoui2006,johnstone2008,johnstonema2011}, it was actually found that the first order correction to the limiting distribution of extreme eigenvalue actually corresponds to a correction of the scaling variable, as in Eq.~(\ref{eq:correction}). One exception concerns the smallest eigenvalue of real Wishart matrices in the soft-edge limit, where the first correction has a more complicated structure~\cite{Ma12}. 

 The main goal of this paper is to provide an explicit computation of this first correction in the hard edge limit and we will show that it has indeed the conjectured form given above in Eq.~(\ref{conjecture_finite_N}). To perform this computation, we will use a method relying on semi-classical OPs, in the spirit of Refs. \cite{Nadal11} and \cite{Chen05,Basor09}. As we will see, our method does not only allow us to compute explicitly the first $1/N$ corrections but provides also a rather straightforward derivation of the expression for the limiting distribution $F_{\infty}(x)$ in terms of the solution of a Painlev\'e III equation, without using Fredholm operators theory but relying instead only on the recurrence relations associated to the (semi-classical) OPs system. Finally, we will also study the first finite $N$ corrections to the limiting distribution of $\lambda_{\min}$ at the soft edge. 

Note that after the results obtained in the present paper were presented in a conference \cite{conf_montevideo}, another independent proof of the conjecture in Eq. (\ref{conjecture_finite_N}) was achieved in Ref. \cite{Bornemann15}, using operator theoretic techniques. More recently, yet another independent proof of this conjecture was given in Ref. \cite{najim2015}.

\section{Summary of main results and outline of the paper}\label{section:Summary}
The distribution of the smallest eigenvalue $\lambda_{\min}=\underset{1 \le i \le N}{\min}\lambda_i$ is given by
\begin{equation}\label{def_FN}
F_N(t)={\rm Prob}(\lambda_{\min} \ge t)=\int_t^{\infty} \!{\rm d}\lambda_1 \int_t^{\infty} \!{\rm d}\lambda_2 ... \int_t^{\infty} \!{\rm d}\lambda_N P_{\rm joint}(\lambda_1,\lambda_2,\cdots,\lambda_N) \;.
\end{equation}
In this paper, we will compute $F_N(t)$ using semi-classical OPs $\{\pi_k(\lambda)\}_{k \in \mathbb{N}}$ which are polynomials of the variable $\lambda$ while $t$ and $a$ are parameters (for the sake of clarity in the notations, this dependence is omitted here):
\begin{eqnarray}\label{eq:def_OP}
\left\{
\begin{array}{l}
  \braket{\pi_k}{\pi_{k'}}=\int_t^{\infty} e^{-\lambda} \lambda^a \pi_k(\lambda) \pi_{k'}(\lambda) {\rm d}\lambda = h_k \delta_{k,k'}, \\
  \\
  \pi_k(\lambda)=\lambda^k + \zeta_k \lambda^{k-1} + ...
\end{array}
\right.
\end{eqnarray}
The cumulative distribution $F_N(t)$ can be expressed, using standard manipulations, in terms of the norms $h_k$'s as
\begin{eqnarray}
F_N(t) = \frac{N!}{Z_N} \prod_{k=0}^{N-1} h_k \;.
\end{eqnarray}
As we will see, the norms $h_k$'s can be computed from the three term recurrence relation satisfied by the OPs
\begin{eqnarray}
\lambda \pi_k=\pi_{k+1}+S_k \pi_k +R_k \pi_{k-1} \;,
\end{eqnarray}
from which we deduce the following important relations:
\begin{eqnarray}\label{important_eq}
&&R_k = \frac{h_k}{h_{k-1}}, \\
&&S_k = - t \partial_t \log h_k + 2k + a + 1 \;,\\
&& \zeta_k = - \sum_{i=0}^{k-1} S_i \;.
\end{eqnarray}
Note that the starting point of our analysis is very similar to the one of Basor and Chen in Ref. \cite{Basor09} but the analysis of the recursion relations is different. In particular, we do not make use of ladder operators techniques, which are heavily used in Ref. \cite{Basor09}. In addition, we provide an asymptotic analysis of this OP system (\ref{eq:def_OP}) for large $N$, beyond the leading order.

In section \ref{section:OP}, we will study the variables $S_k$, $R_k$, $h_k$ and $\zeta_k$. In particular, we will show that $S_k$ and $R_k$
satisfy a coupled set of equations, named Schlesinger equations in the literature on OPs  
\begin{eqnarray}\label{schlesinger_intro}
\left\{
\begin{array}{ccl}
S_k-R_{k+1}+R_{k} &=& t\partial_t S_k,\\
\\
2-S_{k+1}+S_{k} &=& t \dfrac{\partial_t R_{k+1}}{R_{k+1}} \;.
\end{array}
\right.
\end{eqnarray}
This system of equations, together with the initial condition given in Eq. (\ref{CI}) below, determines uniquely the values of $R_k$ and $S_k$ for all values of $k$, as they can be computed by induction. It is however quite difficult to analyze the large $N$ behavior of $R_N$ and $S_N$ using only this set of equations (\ref{schlesinger_intro}). To circumvent this difficulty, it is customary to use another set of relations, called the Laguerre-Freud equations, which we derive here using the method based on Tur\'an determinants, as developed in Ref. \cite{Belmehdi94}. They read
\begin{eqnarray}\label{LF_intro}
\left\{
\begin{array}{rcl}
R_{k+2}-R_k&=&S_{k+1}(2k+4+a+t-S_{k+1})-S_k(2k+a+t-S_k)-2t,\\
\\
S_{k+1}(S_{k+1}-t) &=& R_{k+1}(2k+1+a+t-S_{k+1}-S_k) \\
&-&R_{k+2}(2k+5+a+t-S_{k+2}-S_{k+1}).
\end{array}
\right.
\end{eqnarray}
Some details of this derivation have been relegated in \ref{appendix:Turan}. 

In section \ref{section:PV}, by manipulating these two sets of equations (\ref{schlesinger_intro}) and (\ref{LF_intro}), we show that 
$F_N(t)$, for finite $N$, is related to a special solution of a Painlev\'e V equation [see Eqs. (\ref{def:Hn}, \ref{Painleve5})], thus recovering a previous result of Tracy and Widom. Section \ref{section:PIII} is devoted to the hard edge scaling limit: 
\begin{eqnarray}\label{def_hard_edge}
N \to \infty , \, t \to 0,\,  x=N\, t \in \mathbb{R}^+ \;\;{\rm fixed} \;.
\end{eqnarray}
It is known that the limiting distribution $F_{\infty}(x)$ can be obtained by analyzing the large $N$ limit of this Painlev\'e V equation in Eqs. (\ref{def:Hn}, \ref{Painleve5}), obtained for finite $N$, leading to (\ref{eq:result_TW}) and (\ref{PIII}). As we show in section 5, the finite $N$ corrections are then easily obtained from the Schlesinger equation (\ref{schlesinger_intro}), from which we obtain Eq. (\ref{conjecture_finite_N}). Finally, section \ref{section:PII} is devoted to the soft edge scaling limit, when $a \sim {\cal O}(N)$, for $N$ large.

\section{Semi-classical Orthogonal Polynomials}\label{section:OP}

To study this OP system (\ref{eq:def_OP}) it is useful \cite{Nadal11} to introduce a deformation parameter $\alpha$ and study the following OP system
\begin{eqnarray}\label{eq:OP_deformed}
\left\{
\begin{array}{l}
  \braket{\pi_k}{\pi_{k'}}=\int_t^{\infty} e^{-\alpha\lambda} \lambda^a \pi_k(\lambda) \pi_{k'}(\lambda) {\rm d}\lambda = h_k \delta_{k,k'}, \\
  \\
  \pi_k(\lambda)=\lambda^k + \ldots \;,
\end{array}
\right.
\end{eqnarray}
such that the norms $h_k$'s are defined by
\begin{eqnarray}
h_k &=& \braket{\pi_k}{\pi_{k}}. \label{def_hn}
\end{eqnarray}
As we show here, some useful relations can be obtained by varying $\alpha$. Eventually, we will of course set $\alpha =1$ (\ref{eq:def_OP}). The first polynomials can be computed from (\ref{eq:OP_deformed}) to obtain
\begin{eqnarray}\label{first_polynomials0}
\pi_0(\lambda)&=&1\,,\\
\pi_1(\lambda)&=&\lambda-\frac{1+a}{\alpha}-\frac{e^{-t\alpha}t(t\alpha)^a}{\Gamma(1+a,t\alpha)}\,,\label{first_polynomials1}
\end{eqnarray}
where $\Gamma(\nu,x) = \int_x^\infty y^{\nu-1} e^{-y} {\rm d}y$ is the incomplete gamma function. Obviously, the expression of the OPs $\pi_k$ becomes more and more complicated as $k$ grows. 
The polynomials $\pi_k$ being OPs, they satisfy a three-term recursion relation, which can be obtained as follows. As 
$\lambda \pi_k$ is a polynomial of degree $k+1$, we can expand it on the basis of these OPs. Because $\braket{\pi_{k-2-i}}{\lambda \pi_k}= 0 $ if $i \ge 0$, we can write the three term recurrence relation \cite{Szego75}:
\begin{eqnarray}\label{3term_rec}
\lambda \pi_k=\pi_{k+1}+S_k \pi_k +R_k \pi_{k-1},
\end{eqnarray}
where, by definition
\begin{eqnarray}
S_k h_k &=& \braket{\pi_k}{\lambda\pi_{k}},\\
R_k h_{k-1} &=& \braket{\pi_{k-1}}{\lambda\pi_{k}}.
\end{eqnarray}
From (\ref{first_polynomials0}) and (\ref{first_polynomials1}), we can compute the first terms
\begin{eqnarray}\label{CI0}
\left\{
\begin{array}{ccl}
h_0 &=& \int_t^{\infty} e^{-\alpha\lambda} \lambda^a {\rm d}\lambda = \dfrac{\Gamma(1+a,\alpha t)}{\alpha^{a+1}},\\
\\
S_0 &=& -\partial_\alpha \log{h_0}=\dfrac{1+a}{\alpha}+\dfrac{e^{-t\alpha}t(t\alpha)^a}{\Gamma(1+a,t\alpha)},\\
\\
R_0 &=& 0,\\
\\
\zeta_0&=& 0.
\end{array}
\right.
\end{eqnarray}

\subsection{Schlesinger equations}\label{section:Schlesinger}

In this section, we derive a couple of recursion relations called the Schlesinger equations that couple $R_k$ and $S_k$. We first write
\begin{eqnarray}
R_k h_{k-1} = \braket{\pi_{k-1}}{\lambda\pi_{k}}= \braket{\lambda \pi_{k-1}}{\pi_k} \;.
\end{eqnarray}
Therefore, using Eq. (\ref{3term_rec}) with the substitution $k \to k-1$, we have $\braket{\lambda \pi_{k-1}}{\pi_k} = \braket{\pi_k}{\pi_k} = h_k$ and finally we obtain the standard relation
\begin{eqnarray}\label{R_en_h}
R_k = \frac{h_k}{h_{k-1}}. 
\end{eqnarray}
On the other hand, using the definition of the scalar product in (\ref{eq:OP_deformed}), we have
\begin{eqnarray}\label{Snhn}
S_k h_k = \braket{\pi_k}{\lambda\pi_{k}} = \int_t^{\infty} e^{-\alpha\lambda} \lambda^a \, \lambda \, \pi_{k}^2(\lambda)\, {\rm d}\lambda = -\partial_\alpha \braket{\pi_k}{\pi_{k}} = -\partial_\alpha h_k,
\end{eqnarray}
from which we deduce the relation between $S_k$ and $h_k$
\begin{eqnarray}\label{S_en_h}
S_k = -\partial_\alpha \log{h_k}.
\end{eqnarray}
By combining Eq. (\ref{S_en_h}) and Eq. (\ref{R_en_h}), we obtain straightforwardly
\begin{eqnarray}\label{1_recu}
S_{k+1}-S_{k} = -\partial_\alpha \log{\frac{h_{k+1}}{h_{k}}} =  -\partial_\alpha \log{R_{k+1}}.
\end{eqnarray}
We now study the coefficient $\zeta_k$ of the term of degree $k-1$ in the polynomial $\pi_k$ [see Eq. (\ref{eq:def_OP})]: 
\begin{eqnarray}\label{def_zeta}
\pi_k(\lambda)=\lambda^k + \zeta_k \lambda^{k-1}+...
\end{eqnarray}
Taking the derivative of this equation with respect to (w.r.t.) $\alpha$ we obtain
\begin{eqnarray}\label{eq:deriv_zeta}
\partial_\alpha\pi_k(\lambda)=\partial_\alpha \zeta_k \lambda^{k-1}+ \ldots \;.
\end{eqnarray}
Multiplying both sides of Eq. (\ref{eq:deriv_zeta}) by $\lambda$ and projecting on $\pi_k$ yields
\begin{eqnarray}\label{eq:deriv_zeta2}
\partial_\alpha \zeta_k h_k = \braket{\lambda \partial_\alpha \pi_k}{ \pi_{k}}.
\end{eqnarray}
Besides, by looking at the term of degree $k$, i.e. $\propto \lambda^k$, in Eq. (\ref{3term_rec}) we find a recursion relation between $S_k$ and $\zeta_k$
\begin{eqnarray}\label{zeta_S}
\zeta_k = \zeta_{k+1}+S_k \;,
\end{eqnarray}
which can be solved for $\zeta_k$, using the initial condition (\ref{CI0}), to get
\begin{eqnarray}\label{zeta_k_S}
\zeta_k = -\sum_{i=0}^{k-1} S_i.
\end{eqnarray}
Furthermore, by differentiating Eq. (\ref{Snhn}) w.r.t. to $\alpha$ we have
\begin{eqnarray}
\partial_\alpha (S_k h_k) &=& - \int_t^\infty e^{-\alpha \lambda} \lambda^\alpha \, \lambda^2 \, \pi_k^2(\lambda) \, {\rm d}\lambda + 2 \int_t^\infty e^{-\alpha \lambda} \lambda^\alpha \, \lambda \,\partial_\alpha \pi_k(\lambda) \pi_k(\lambda) \, {\rm d }\lambda \nonumber \\
&=& -\braket{\lambda \pi_k}{\lambda\pi_{k}} + 2 \braket{\lambda \partial_\alpha \pi_k}{\pi_{k}} \;,
\end{eqnarray}
where, in the second line, we have simply used the definition of the scalar product in (\ref{eq:def_OP}). Using the three-term recurrence relation (\ref{3term_rec}) to rewrite the first term and Eq. (\ref{eq:deriv_zeta2}) to rewrite the second one, we have
\begin{eqnarray}\label{deriv_prod}
\partial_\alpha (S_{k} h_k)=  -h_{k+1} -S^2_k h_k -R_k^2 h_{k-1} +2 \partial_\alpha \zeta_k h_k \;.
\end{eqnarray}
We can also write
\begin{eqnarray}\label{id_prod}
\partial_\alpha (S_{k} h_k)= (\partial_\alpha S_{k}) h_k+S_k (\partial_\alpha h_k) \;.
\end{eqnarray}
By replacing the left hand side of Eq. (\ref{deriv_prod}) by Eq. (\ref{id_prod}), dividing the resulting equation by $h_k$ and using 
Eqs. (\ref{R_en_h}) and (\ref{S_en_h}), we obtain a last recursion relation between $R_k$, $S_k$ and $\zeta_k$
\begin{eqnarray}\label{2_recu}
R_{k+1} +R_k = -\partial_\alpha S_{k} +2 \partial_\alpha \zeta_k \;.
\end{eqnarray}
Combining Eq. (\ref{zeta_S}) and (\ref{2_recu}), we have
\begin{eqnarray}
R_{k+1}+R_{k} = \partial_\alpha \zeta_{k+1}+ \partial_\alpha \zeta_k.
\end{eqnarray}
And finally, with the initial condition $R_0 = \zeta_0 = 0$ in Eq. (\ref{CI0}) we obtain
\begin{eqnarray}\label{eq:rel_zeta_R}
\partial_\alpha \zeta_k = R_k \;.
\end{eqnarray}
Therefore substituting this relation in Eq. (\ref{2_recu}) we finally obtain a closed system of two coupled recursion relations between $S_k$ and $R_k$
\begin{eqnarray}\label{recu_derivalpha}
\left\{
\begin{array}{ccl}
R_{k+1}-R_{k} &=& -\partial_\alpha S_k,\\
\\
S_{k+1}-S_{k} &=& -\partial_\alpha \log{R_{k+1}} \;,
\end{array}
\right.
\end{eqnarray}
where the second equation was previously obtained in Eq. (\ref{1_recu}).

Our goal now is to relate $\alpha$-derivatives to $t$-derivatives, i.e., find a relation between $\partial_\alpha h_k$ and $\partial_t h_k$. Differentiating Eq. (\ref{def_hn}) w.r.t. $t$, using that $\braket{\partial_t \pi_k}{\pi_{k}}=0$ as $\partial_t \pi_k$ is a polynomial of degree $k-1$ (\ref{eq:def_OP}), we have
\begin{eqnarray}\label{dot_hn}
\partial_t h_k=-e^{-\alpha t}t^{a} \pi_k^2(t).
\end{eqnarray}
We now start from the expression of $\partial_\alpha h_k$ given in Eq. (\ref{Snhn}) and use integration by parts to obtain
\begin{eqnarray}
-\partial_\alpha h_k&=& \int_t^\infty{\rm d}\lambda \pi_k^2(\lambda)e^{-\alpha\lambda}\lambda^{a+1}, \nonumber \\
&=&\big[-\frac{1}{\alpha} e^{-\alpha\lambda}\lambda^{a+1} \pi_k^2(\lambda)\big]^{\lambda=\infty}_{\lambda=t}\\
&+&\frac{1}{\alpha}\int_t^\infty{\rm d}\lambda e^{-\alpha\lambda}\lambda^{a} \left((a+1) \pi_k^2(\lambda) +  2 \lambda
 \pi_k(\lambda) \partial_\lambda\pi_k(\lambda)\right),\nonumber \\
 &=&\frac{1}{\alpha} (-t\partial_t h_k+(a+1)h_k+2\braket{ \pi_k}{\lambda\partial_\lambda\pi_{k}}) \;,
\end{eqnarray}
where we have used Eq. (\ref{dot_hn}). We can easily calculate the last scalar product
\begin{eqnarray}
\braket{ \pi_k}{\lambda\partial_\lambda\pi_{k}}=\braket{\pi_k}{(k \lambda^k + ...)}=k h_k \;,
\end{eqnarray}
to obtain the desired relation between $\partial_\alpha h_k$ and $\partial_t h_k$
\begin{eqnarray}\label{deriv_alpha_t_h}
\alpha \partial_\alpha h_k=t\partial_t h_k-(2k+a+1)h_k.
\end{eqnarray}
With this relation (\ref{deriv_alpha_t_h}), it is then straightforward to relate $\partial_\alpha R_k$ and $\partial_\alpha S_k$ to $\partial_t R_k$ and $\partial_t S_k$, using  (\ref{R_en_h}) and (\ref{S_en_h}). Note that from now on, we set $\alpha =1$
\begin{eqnarray}\label{S_en_h_sans_alpha}
S_k = -t \partial_t \log{h_k}+(2k+a+1).
\end{eqnarray}
We finally obtain, using Eq. (\ref{recu_derivalpha}), the so-called Schlesinger equations \cite{Magnus95}
\begin{eqnarray}
\left\{
\begin{array}{ccl}\label{recurrence_dotS}
S_k-R_{k+1}+R_{k} &=& t\partial_t S_k,\\
\\
2-S_{k+1}+S_{k} &=& t \partial_t \log R_{k+1}.
\end{array}
\right.
\end{eqnarray}
Note that, for $\alpha =1$, the initial condition (\ref{CI0}) reads
\begin{eqnarray}\label{CI}
\left\{
\begin{array}{ccl}
h_0 &=& \int_t^{\infty} e^{-\lambda} \lambda^a{\rm d}\lambda = \Gamma(1+a,t),\\
\\
S_0 &=& -t \partial_t \log{h_0} +a+1 = \frac{e^{-t}t^{a+1}}{\Gamma(1+a,t)}+a+1,\\
\\
R_0 &=& 0,\\
\\
\zeta_0&=& 0.
\end{array}
\right.
\end{eqnarray}
Using the Schlesinger equations (\ref{recurrence_dotS}) and this initial condition (\ref{CI}), we can compute step by step all the terms for arbitrary $k$.

We end up this section by providing a useful relation between $\zeta_k$ [see Eq. (\ref{def_zeta})] and the cumulative distribution of the smallest eigenvalue $F_N(t)={\rm Prob}(\underset{1 \le i \le N}{\min}\lambda_i \ge t)$. One has
\begin{equation}\label{F_k_fonction_de_hn}
F_N(t)=\int_t^{\infty} \!{\rm d}\lambda_1 \int_t^{\infty} \!{\rm d}\lambda_2 ... \int_t^{\infty} \!{\rm d}\lambda_N \, P_{\rm joint}(\lambda_1,\lambda_2,\cdots,\lambda_N) = \frac{N!}{Z_N} \prod_{k=0}^{N-1} h_k,
\end{equation}
where the last equality is obtained by using the classical tricks of replacing the Vandermonde determinant by the determinant built from the OPs $\pi_k$'s (\ref{eq:def_OP}) and then use the Cauchy-Binet formula \cite{Mehta91,Forrester10}. Using Eq. (\ref{zeta_k_S}) with (\ref{S_en_h_sans_alpha}) as well as (\ref{deriv_alpha_t_h}) , we can write
\begin{eqnarray}
\zeta_N=-N(N+a)+t\partial_t\log\left(\prod_{k=0}^{N-1} h_k \right).
\end{eqnarray}
And therefore, from the expression of $F_N(t)$ given in Eq. (\ref{F_k_fonction_de_hn}), we have
\begin{eqnarray}\label{lien_zeta_F}
\zeta_N=-N(N+a)+t\partial_t\log \left(F_N(t)\right) \;.
\end{eqnarray}
This expression thus provides a link between $\zeta_N$, which is the first non trivial coefficient of the OP's [see Eq.  (\ref{def_zeta})] and the cumulative distribution of the smallest eigenvalue. 

\subsection{Laguerre-Freud equations}\label{section:LF}

In this section, we derive another set of recursion relations between the coefficients $R_k$'s and $S_k$'s, the so-called Laguerre-Freud equations, following the procedure used by Belmehdi and Ronveaux in \cite{Belmehdi94}. To derive these equations, we start by searching two functions $\Psi$ and $\Phi$, which are polynomials in $\lambda$, satisfying for any polynomial~$p$
\begin{eqnarray}\label{def_phi_psi}
\braket{\Psi}{p}=\braket{\Phi}{p'} \;,
\end{eqnarray}
where the polynomials $\Phi$ and $\Psi$ may depend explicitly on the parameters $t$ and $a$ and $p'(\lambda)\equiv \partial_\lambda p(\lambda) $. Denoting by $w(\lambda)=e^{-\lambda} \lambda^a$ the weight associate to the scalar product in Eq. (\ref{eq:def_OP}) and using an integration by parts, this relation (\ref{def_phi_psi}) is satisfied provided that
\begin{eqnarray}
\Psi w+ (\Phi w)' = 0 \hspace{0.2cm} {\rm and} \hspace{0.2cm} \Phi(\lambda=t)=0.
\end{eqnarray}
We find that the simplest non trivial solution to this equation is given by
\begin{eqnarray}
\left\{
\begin{array}{ccl}
\Psi(\lambda)&=&\lambda^2-(2+a+t)\lambda+t(1+a),\\
\\
\Phi(\lambda)&=&\lambda^2-t \lambda \;.
\end{array}
\right.
\end{eqnarray}
Then, using this property (\ref{def_phi_psi}) for $p=\pi_k^2$ and for $p=\pi_k\pi_{k+1}$, we can write after expansion, using the three-term 
recursion relation (\ref{3term_rec})
\begin{eqnarray}
\left\{
\begin{array}{ccl}
I_{2,k}-(2+a+t)I_{1,k}+t(1+a)I_{0,k}&=& 2(J_{2,k}-tJ_{1,k}),\\
\\
K_{2,k}-(2+a+t)K_{1,k}+t(1+a)K_{0,k}&=& L_{2,k}-tL_{1,k},
\end{array}
\right.
\end{eqnarray}
where we have introduced the four family of integrals 
\begin{eqnarray}
\left\{
\begin{array}{ccl}
I_{m,k} &=& \braket{\lambda^m}{\pi_{k}^2},\\
\\
J_{m,k} &=& \braket{\lambda^m}{\pi_{k}\pi'_{k}},\\
\\
K_{m,k} &=& \braket{\lambda^m}{\pi_{k}\pi_{k+1}},\\
\\
L_{m,k} &=& \braket{\lambda^m}{(\pi_{k+1}\pi_k)'}.
\end{array}
\right.
\end{eqnarray}
These integrals are all calculated (for $m=0, 1$ and $2$) in \ref{appendix:Turan}. 
Using these expressions together with Eq. (\ref{R_en_h}), we find the two relations
\begin{eqnarray}
\left\{
\begin{array}{rcl}\label{LF01}
&&R_{k+1}+R_k+S_k(S_k-a-t-2k-2)+t(2k+1+a) = 2 \,\displaystyle{\sum_{i=0}^{k-1} }S_i, \\
\\
&&(S_{k+1}+S_k-3-a-t-2k)R_{k+1} = 2 \,\displaystyle{\sum_{i=1}^k R_i+\sum_{i=0}^k S_i^2 - t \sum_{i=0}^k S_i}.
\end{array}
\right.
\end{eqnarray}
We can rewrite these equations by subtracting rank $k$ to the rank $k+1$ and find the two Laguerre-Freud recurrence equations (which are here of order 2):
\begin{eqnarray}\label{laguerre-freud}
\left\{
\begin{array}{rcl}
R_{k+2}-R_k&=&S_{k+1}(2k+4+a+t-S_{k+1})-S_k(2k+a+t-S_k)-2t,\\
\\
S_{k+1}(S_{k+1}-t) &=& R_{k+1}(2k+1+a+t-S_{k+1}-S_k)\\
&-&R_{k+2}(2k+5+a+t-S_{k+2}-S_{k+1}).
\end{array}
\right.
\end{eqnarray}
%and 
%\begin{eqnarray}
%2\omega_k =-t \partial_t S_k+ (t-S_k)\theta_k \label{omega_k_det_theta}.
%\end{eqnarray}
As we show below these two sets of equations (\ref{recurrence_dotS}) and (\ref{laguerre-freud}) allow us to (i) derive the connection to the Painlev\'e equation and (ii) perform the asymptotic analysis of these coefficients for large $N$.

\section{Painlev\'e V equation for finite $N$}\label{section:PV}

In this section, we proceed to the derivation of the Painlev\'e V equation, following the method of Ref. \cite{Basor09}. First, it is useful to introduce the quantities~\cite{Forrester07}
\begin{eqnarray}
\left\{
\begin{array}{rcl}\label{def_omega_txt}
\theta_k&=&2k+1+a-S_k,\\
\\
\omega_k&=&-R_k-\zeta_k. 
\end{array}
\right.
\end{eqnarray}
Manipulating the Laguerre-Freud equations (\ref{laguerre-freud}) we can prove (see \ref{appendix:LF})
\begin{eqnarray}\label{LF02_bonne_limite}
\omega_k^2 -\theta_k (\theta_k-a-2k+t) \omega_k -\theta_k (kt(k+a)+(\theta_k+t)\zeta_k) = 0 \;,
\end{eqnarray}
which is a simple algebraic relation (and not a recursion relation) between the different variables $\omega_k, \theta_k$ and $\zeta_k$. 
Using the previous relations with the index $k=N$, we find from Eq. (\ref{lien_zeta_F}),
\begin{eqnarray}\label{def:Hn}
H_N=t \partial_t \log(F_N(t)) =N(N+a)+ \zeta_N.
\end{eqnarray}
Summing up the first Schlesinger equation (\ref{recurrence_dotS}) from $k=0$ to $k=N-1$, we find
\begin{eqnarray}
\sum_{k=0}^{N-1} S_k + \sum_{k=0}^{N-1}(R_k-R_{k-1})= t \partial_t \sum_{k=0}^{N-1} S_k
\end{eqnarray}
which can be simplified using (\ref{zeta_k_S}), (\ref{CI0}) and (\ref{def_omega_txt}) to yield
\begin{eqnarray}\label{zeta:omegan}
-t \partial_t \zeta_N = -R_N-\zeta_N =\omega_N .
\end{eqnarray}
On the other hand, using (\ref{def:Hn}) and (\ref{zeta:omegan}), we have
\begin{eqnarray}\label{Hn_to_Rn}
t \partial_t H_N = - \omega_N = \zeta_N+R_N =  H_N -N(N+a)+R_N \;.
\end{eqnarray}
Taking a derivative of this equation w.r.t. $t$, we find
\begin{eqnarray}\label{partialRn:H''}
\partial_t R_N = t \,\partial_t^2 H_N.
\end{eqnarray}
From the second Schlesinger equation (\ref{recurrence_dotS}), we find
\begin{eqnarray}
t \partial_t R_N &=& R_N (2-S_N-S_{N-1}) = R_N (\theta_N-\theta_{N-1}) \;,
\end{eqnarray}
where we have used (\ref{def_omega_txt}). Using the relation derived in \ref{appendix:LF} in Eq. (\ref{theta_km}), we can rewrite this recursion equation as a simple algebraic relation
\begin{eqnarray}\label{partialR_k:omegaRnthetan}
t \partial_t R_N &=& R_N \theta_N-\frac{\omega_N^2}{\theta_N} \;.
\end{eqnarray}
Therefore, combining (\ref{partialRn:H''}) and (\ref{partialR_k:omegaRnthetan}), we find
\begin{eqnarray}\label{H'':omegaRnthetan}
t^2 \partial_t^2 H_N &=& R_N \theta_N-\frac{\omega_N^2}{\theta_N} \;.
\end{eqnarray}
On the other hand, from our Eq. (\ref{LF02_bonne_limite}) in which we inject the definition of $\omega_N$ (\ref{def_omega_txt}) to eliminate $\zeta_N$
\begin{eqnarray}\label{LF_modif_kozeta}
N(N+a)t-(2N+a)\omega_N-t R_N &=& R_N \theta_N+\frac{\omega_N^2}{\theta_N}.
\end{eqnarray}
Summing and subtracting the last two equations (\ref{H'':omegaRnthetan}) and (\ref{LF_modif_kozeta}), we obtain
\begin{eqnarray}
2\frac{\omega_N^2}{\theta_N} &=& N(N+a)t-(2N+a)\omega_N-t R_N -t^2 \partial_t^2 H_N, \label{multi1}\\
2 R_N \theta_N &=&  N(N+a)t-(2N+a)\omega_N-t R_N +t^2 \partial_t^2 H_N. \label{multi2}
\end{eqnarray}
Multiplying these two equations (\ref{multi1}) and (\ref{multi2}) together, we find
\begin{eqnarray}
4 R_N \omega_N^2 &=& \left(N(N+a)t-(2N+a)\omega_N-t R_N \right)^2- \left(t^2 \partial_t^2 H_N\right)^2.
\end{eqnarray}
Finally, by using (\ref{Hn_to_Rn}), we eliminate $\omega_N$ and $R_N$ (by expressing them in terms of $H_N$ and $\partial_t H_N$) and find the equation 
\begin{equation}\label{Painleve5}
\left(t \partial_t^2 H_N\right)^2 = 4\left( \partial_t H_N \right)^2 \left( H_N -N(N+a) -t \partial_t H_N \right) + \left((2N+a-t)\partial_t H_N +H_N\right)^2 \;,
\end{equation}
which is a Painlev\'e V equation in the Jimbo-Miwa-Okamoto $\sigma$ form \cite{Jimbo81,Okamoto82}. Note that this equation coincides exactly with the equation first found by Tracy and Widom in \cite{Tracy94b}.

\section{Large $N$ asymptotic limit at the hard edge: Painlev\'e III and first correction}\label{section:PIII}

In this section, we study the behavior of the quantities $h_N$, $R_N$, $S_N$ and $\zeta_N$ in the hard edge limit (\ref{def_hard_edge}), which, in the language of OPs, corresponds to a double scaling limit. Of course, as we are eventually interested in the study of the cumulative distribution of the smallest eigenvalue $F_N(t)$, we could perform this asymptotic analysis directly on the Painlev\'e V equation (\ref{Painleve5}), as done by Tracy and Widom in their original study of GUE \cite{TW94a}. But it turns out to be much more convenient, especially to extract the $1/N$ corrections, to perform this asymptotic analysis on the Schlesinger (\ref{recurrence_dotS}) and Laguerre-Freud (\ref{laguerre-freud}) equations.

To understand the structure of the coefficients $h_N$, $R_N$, $S_N$ and $\zeta_N$ in this double scaling limit (\ref{def_hard_edge}), it is useful to study their behavior for small $t$ (keeping $N$ fixed). For $t=0$, the OPs $\pi_k$'s in Eq. (\ref{eq:def_OP}) can be expressed in terms of the generalized Laguerre polynomials  
\begin{eqnarray}\label{Laguerre:def}
L_{k}^{(a)}(x)= \frac{\Gamma(a+k+1)}{k!} \sum_{i=0}^k \binom{k}{i} \frac{(-x)^i}{\Gamma(a+i+1)}\,.
\end{eqnarray}
Hence, thanks to standard properties of Laguerre polynomials, we easily obtain these coefficients, for $t=0$, 
using (\ref{def_hn}), (\ref{R_en_h}) and (\ref{S_en_h_sans_alpha}) as:
\begin{eqnarray}
\left\{
\begin{array}{ccl}\label{polym0}
\left.  \pi_k(\lambda) \right|_{t=0}  &=& L_k^{(a)}(\lambda) k! (-1)^k,\\
\\
\left.  h_k \right|_{t=0} &=&\Gamma(k+a+1)k!,\\
\\
\left.  R_k \right|_{t=0} &=&k(k+a),\\
\\
\left.  S_k \right|_{t=0} &=&2k+a+1,\\
\\
\left.  \zeta_k \right|_{t=0} &=&-k(k+a).
\end{array}
\right.
\end{eqnarray}
We now take the index $k=N$. When $t$ is closed to 0, we can use Eq. (\ref{dot_hn}) and find
\begin{equation}\label{Eq:dot_h}
\partial_t h_N = -t^a e^{- t}\pi_N(t)^2 = -t^a  L_N^{(a)}(0)^2 N!^2 +o(t^a)= -t^a \frac{[\Gamma(N+a+1)]^2}{[\Gamma(a+1)]^2} +o(t^a) \;.
\end{equation}
Using (\ref{polym0}), we can integrate and find the first correction
\begin{eqnarray}
h_N=\Gamma(N+a+1)N!-\frac{t^{a+1}[\Gamma(N+a+1)]^2}{\Gamma(a+2) \Gamma(a+1)}+o(t^{a+1}).
\end{eqnarray}
We are interested in the scaling behavior when $N$ goes to infinity, $t$ goes to $0$ with $x=Nt$ finite (\ref{def_hard_edge}). In this double scaling regime, the expansion above reads
\begin{eqnarray}\label{hn_small_x}
h_N &=&\Gamma(N+a+1)N!\left[1-\frac{1}{N} \left(\frac{x^{a+1}}{\Gamma(a+2)\Gamma(a+1)}+ o(x^{a+1})\right) + o\left(\frac{1}{N}\right)\right].
\end{eqnarray}
This suggests the ansatz
\begin{eqnarray}\label{ansatz}
h_N=\Gamma(N+a+1)N!\left[1+\frac{1}{N}f(x= N\,t) + o\left(\frac{1}{N}\right)\right],
\end{eqnarray}
where the function $f$ has thus the small $x$ expansion, read from Eq.~(\ref{hn_small_x}) 
\begin{eqnarray}\label{f_dvpt0}
f(x) &=& -\frac{x^{a+1}}{\Gamma(a+2)\Gamma(a+1)} + o(x^{a+1}).
\end{eqnarray}
We can introduce the ansatz (\ref{ansatz}) in the relations (\ref{R_en_h}) and (\ref{S_en_h_sans_alpha}) which give $R_N$ and $S_N$ in terms of the function $f$. We obtain
\begin{eqnarray}\label{ansatz_R_S}
\left\{
\begin{array}{ccl}
R_N&=&N(N+a)+x f'(x) -f(x) + o\left(1\right),\\
\\
S_N&=&2N+a+1-\dfrac{1}{N}x f'(x) + o\left(\dfrac{1}{N}\right).
\end{array}
\right.
\end{eqnarray}
The first non trivial terms in Eq. (\ref{ansatz_R_S}) $xf'(x) - f(x)$ for $R_N$ and $-(1/N) x \,f'(x)$ for $S_N$ are necessary to compute the leading order of the cumulative distribution, $F_\infty(x)$. To compute the first $1/N$ correction to the limiting distribution, we need to expand $R_N$ and $S_N$ in Eq. (\ref{ansatz_R_S}) to the next order in $1/N$. One actually expects the following expansion 
\begin{eqnarray}\label{R_S_expansion}
\left\{
\begin{array}{ccl}
R_N&=&N(N+a)+\displaystyle \sum_{i=0}^j r_i(x)N^{-i} + o\left(N^{-j}\right),\\
\\
S_N&=&2N+a+1+\displaystyle \sum_{i=1}^j s_i(x)N^{-i} + o\left(N^{-j}\right) \;,
\end{array}
\right.
\end{eqnarray}
where the first term in this expansion are given in (\ref{ansatz_R_S}), i.e., $r_0(x) = xf'(x)-f(x)$ and $s_1(x) = - xf'(x)$. Besides, using the exact relation $\partial_t h_N = -t^a e^{- t}\pi_N(t)^2$ [see Eq. (\ref{Eq:dot_h})], one can show that $r_i(x) = o(x)$ as well as $s_i(x)=o(x)$ when $x \to 0$. One can check, in principle, the validity of this expansion (\ref{R_S_expansion}) order by order in powers of $1/N$ by injecting it in the Schlesinger equations (\ref{recurrence_dotS}). Proving it rigorously to arbitrary order $j$ is however a hard task. However, here, we only need this asymptotic expansion up to order ${\cal O}(1/N^2)$, which can be obtained explicitly as follows. To compute the second correction, we truncate the expansion (\ref{R_S_expansion}) up to the $N^{-2}$ order and write
\begin{eqnarray}\label{R_S_expansion2}
\left\{
\begin{array}{ccl}
R_N&=&N(N+a)+x f'(x)-f(x)+\dfrac{r_1(x)}{N}+\dfrac{r_2(x)}{N^2}+o(N^{-2})\,,
\\
\\
S_N&=&2N+a+1-\dfrac{x f'(x)}{N}+\dfrac{s_2(x)}{N^2}+o(N^{-2})\,.
\end{array}
\right.
\end{eqnarray}
In the hard edge limit $N \to \infty$, $t \to 0$, keeping $x=N t$ fixed, we can obtain the expansion of $R_{N+1}$ and $S_{N+1}$ at the same order by replacing $N$ by $N+1$ and using $(N+1)t = x(1 + 1/N)$ in (\ref{R_S_expansion2})
\begin{eqnarray}\label{R_S_expansion}
\left\{
\begin{array}{ccl}
R_{N+1}&=&(N+1)(N+1+a)+x f'(x)-f(x)+\frac{1}{N}(r_1(x)+x^2 f''(x))\\
&&\!\!\!+\frac{1}{N^2}(r_2(x)+x r_1'(x)-r_1(x)+\frac{1}{2}x^2 f''(x)+\frac{1}{2}x^3 f'''(x))+o(N^{-2})\,,
\\
\\
S_{N+1}&=&2N+a+3-\frac{1}{N}x f'(x)+\frac{1}{N^2}(s_2(x)-xf'(x)-x^2f''(x))+o(N^{-2})\,.
\end{array}
\right.
\end{eqnarray}
By injecting these two expansions into the Schlesinger equation (\ref{recurrence_dotS}), we find (using $t \partial_t = x \partial_x$) at the first non trivial order (${\cal O}(N^{-2})$ for the first Schlesinger equation and at ${\cal O}(N^{-3})$ for the second)
\begin{eqnarray}\label{R_S_expansion2}
\left\{
\begin{array}{ccl}
s_2(x)-xs'_2(x)+r_1(x)-xr_1'(x)&=&\frac{1}{2}(x^2 f''(x)+x^3 f'''(x))\,,\\
\\
2s_2(x)-xs'_2(x)-xr_1'(x)&=&-a x^2 f''(x)+\frac{1}{2}x^3 f'''(x))\,,
\end{array}
\right.
\end{eqnarray}
which can be solved as follows. First, by subtracting the first equation of (\ref{R_S_expansion2}) to the second one, one obtains
\begin{eqnarray}\label{rel_s2_r1}
s_2(x) = r_1(x) - \left(a+\frac{1}{2}\right)x^2 f''(x) \;.
\end{eqnarray} 
By injecting this relation (\ref{rel_s2_r1}) in the first equation of (\ref{R_S_expansion2}), one finds that $r_1$ satisfies the following equation
\begin{eqnarray}\label{eq_r1_only}
r_1(x) - x r_1'(x) = - \frac{a}{2} \left(x^2 f''(x) + x^3 f'''(x)\right) \;,
\end{eqnarray}
which can be solved, using that $r_1(x) = o(x)$ as $x \to 0$, yielding
\begin{eqnarray}\label{expr_r1}
r_1(x) = \frac{a}{2} x^2 f''(x) \;.
\end{eqnarray}
Consequently, from Eq. (\ref{rel_s2_r1}) one has
\begin{eqnarray}\label{expr_s2}
s_2(x) = - \frac{a+1}{2} x^2 f''(x) \;.
\end{eqnarray}
We finally find the two first terms of the expansion of $R_N$ and $S_N$, for large $N$, as
\begin{eqnarray}\label{expansion_second_order}
\left\{
\begin{array}{rcl}
R_N&=&N(N+a) + xf'(x)-f(x) + \dfrac{a}{2N}x^2 f''(x)+o\left(\dfrac{1}{N}\right),\\
\\
S_N&=&2N+a+1 - \dfrac{1}{N}xf'(x)-\dfrac{a+1}{2N^2} x^2 f''(x)+o\left(\dfrac{1}{N^2}\right).
\end{array}
\right.
\end{eqnarray} 
From the expansion of $S_N$ in Eq. (\ref{expansion_second_order}), we compute $\zeta_N$ in the double scaling limit (\ref{def_hard_edge}), using Eq.~(\ref{zeta_S}), up to the second non-trivial order for large $N$
\begin{eqnarray} \label{ansatz_zeta}
\zeta_N&=&-(N + a) N + f(x) + \frac{a}{2N} x f'(x) + o\left(\frac{1}{N}\right) \;.
\end{eqnarray}
Finally, from the expression of $\zeta_N$ we obtain the large $N$ expansion of $F_N(t)$ in the hard edge limit (\ref{def_hard_edge}), using 
Eq. (\ref{lien_zeta_F}), which is given by
\begin{eqnarray}\label{Dvpt_F0}
x\partial_x\log\left(F_N\left(\frac{x}{N}\right) \right) = f(x) + \frac{a}{2N} x f'(x) + o\left(\frac{1}{N}\right).
\end{eqnarray}
Using the initial condition $F_N(0)=1$ and (\ref{f_dvpt0}) we can rewrite this equation as
\begin{eqnarray}\label{Dvpt_F}
F_N\left(\frac{x}{N}\right) = \exp{ \left( \int_0^x \frac{f(u)}{u} {\rm d}u + \frac{a}{2N} f(x) + o\left(\frac{1}{N}\right) \right) }.
\end{eqnarray}
Expanding Eq. (\ref{Dvpt_F}) up to first order in $1/N$, one obtains finally
\begin{eqnarray}\label{Finfty2}
F_N\left(\frac{x}{N}\right) = F_\infty(x) + \frac{a}{2N} x F'_\infty(x) +  o\!\left(\frac{1}{N}\right) \;,
\end{eqnarray}
where $F_{\infty}(x)$ is given by 
\begin{eqnarray}\label{Finfty(x)}
\underset{N \to \infty}{\lim}F_N\left(\frac{x}{N}\right) = F_\infty(x)\,,\hspace{0.7cm} F_\infty(x)=\exp\left( \int_0^x \frac{f(u)}{u} {\rm d}u\right) \;.
\end{eqnarray}
Hence we easily obtain the functional form of the $1/N$ correction as given in Eq.~(\ref{Finfty2}), fully consistent with the conjecture in (\ref{conjecture_finite_N}) made in \cite{Edelman14}. However, at this stage, we still need to find the equation satisfied by the function $f$, which enters in the definition of $F_{\infty}(x)$ in Eq. (\ref{Finfty(x)}). To obtain this equation, we expand 
$\theta_N$ and $\omega_N$ in Eq.~(\ref{def_omega_txt}), which is easily done from the expansions of $R_N$, $S_N$ and $\zeta_N$ in (\ref{ansatz_R_S}) and (\ref{ansatz_zeta}) to yield 
\begin{eqnarray}\label{dvpt_omega_theta}
\left\{
\begin{array}{rcl}
\theta_N&=& \dfrac{1}{N}xf'(x)+\dfrac{a+1}{2N^2} x^2 f''(x)+o\left(\dfrac{1}{N^2}\right),\\
\\
\omega_N&=& - xf'(x) - \dfrac{a}{2N}(x^2 f''(x)+x f'(x)) + o\left(\dfrac{1}{N}\right). 
\end{array}
\right.
\end{eqnarray}
Finally, by injecting these expansions (\ref{ansatz_zeta}) and (\ref{dvpt_omega_theta}) in (\ref{LF02_bonne_limite}) with $t=x/N$, we obtain, by canceling the first term, of order ${\cal O}(N^{-2})$, in Eq. (\ref{LF02_bonne_limite}) that $f$ satisfies a Painlev\'e III equation
\begin{eqnarray}
(xf'')^2+4f'(1+f')(xf'-f)=(a f')^2, \label{PainleveIII}
\end{eqnarray}
with the small argument behavior in Eq. (\ref{f_dvpt0}). This result coincides with the one obtained, by a quite different method, by Tracy and Widom \cite{Tracy94} (note the correspondence $\sigma(s)=-f(s/4)$, where $\sigma(s)$ is the notation used in \cite{Tracy94}). Note that for integer values of $a$, $f(x)$ can be written explicitly in terms Bessel functions \cite{Forrester07} [see Eq. (\ref{fdetBessel})]. These results in Eqs. (\ref{Finfty2}), (\ref{Finfty(x)}) and (\ref{PainleveIII}) yield the results announced in the introduction in Eqs. (\ref{eq:result_TW}), (\ref{PIII}) and (\ref{conjecture_finite_N}). Finally, in \ref{appendix:Numeric} we present a comparison between numerical simulations of Wishart matrices of size $N=50$ and the asymptotic formula in Eq. (\ref{Finfty2}) describing the first $1/N$ correction.

\section{Large $N$ asymptotic limit at the soft edge: Painlev\'e II and first correction}\label{section:PII}

We now turn to the analysis of the PDF of the smallest eigenvalue $\lambda_{\min}$ in the case where $a \sim {\cal O}(N)$, and we set $a = \alpha \, N$. In this case, the density of eigenvalues has a single support on $[N x_-, N x_+]$ [see Eq. (\ref{density_MP})], with $x_{\pm} = (\sqrt{1+\alpha} \pm 1)^2$ with a soft edge at both extremities (see Fig. \ref{fig:Wishart_soft_vs_hard}). Therefore, one expects that $\lambda_{\min}$ will be close to $N x_-$, while its fluctuations, of order ${\cal O}(N^{1/3})$, are governed by the Tracy-Widom distribution for $\beta = 2$. In the soft edge limit, the large $N$ analysis of $F_N(t)$ in Eq. (\ref{def_FN}) is more conveniently done directly on the Painlev\'e V equation (\ref{Painleve5}) \cite{Baker98}. Following this route, one can indeed show \cite{Baker98} that, for large $N$ 
\begin{eqnarray}\label{baker_forrester}
\lambda_{\min} = N x_- - \frac{N^{1/3}}{m}\chi + {o}(N^{1/3})\;,
\end{eqnarray}
where $m$ is given by \cite{Baker98}
\begin{eqnarray}\label{def_m}
m=\frac{(1+\alpha)^{1/6}}{(\sqrt{1+\alpha}-1)^{4/3}} \;,
\end{eqnarray}
and where ${\chi}$ is distributed according to the Tracy-Widom distribution ${\cal F}_2$, i.e. $\Pr[\chi \leq s] = {\cal F}_2(s)$ where ${\cal F}_2(s)$ is given by \cite{TW94a}
\begin{eqnarray}\label{TW2}
{\cal F}_2(s) = \exp{\left(-\int_s^\infty (x-s) q^2(x) \, {\rm d}x \right)} \;.
\end{eqnarray}
Here $q(x)$ is the so-called Hatings-McLeod solution of the Painlev\'e II equation
\begin{eqnarray}\label{PII}
q''(x) = x\,q(x) + 2 q^3(x) \;, {\rm with} \; q(x) \sim {\rm Ai}(x) \;, \; {\rm for}\; x \to \infty \;, 
\end{eqnarray}
where ${\rm Ai}(x)$ is the Airy function. The result in Eq. (\ref{baker_forrester}) can be equivalently written as
\begin{eqnarray}\label{tilde_f0}
F_N(t) = \tilde f_0\left(m \frac{N\, x_- -t}{N^{1/3}}\right) + {o}(1) \;,
\end{eqnarray}
which implies 
\begin{eqnarray}\label{HN_0}
H_N(t) = t \partial_t \log F_N(t) &=& -(m x_-) \tilde h_0(x) N^{2/3} + o(N^{2/3}) \;, \; x = m \frac{N\, x_- -t}{N^{1/3}}.
\end{eqnarray}
with $\tilde h_0(x)=\tilde f_0'(x)/f_0(x)$ and where $\tilde f_0 = {\cal F}_2$, and where the tilde refers to the soft edge scaling limit. By injecting this form (\ref{HN_0}) into the Painlev\'e V equation satisfied by $H_N(t)$ (\ref{Painleve5}) one finds that $\tilde h_0$ satisfies the following equation
\begin{eqnarray}\label{PII_bis}
\left(\tilde h_0''(x)\right)^2+ 4 \tilde h'_0(x) \left[\left(\tilde h'_0(x)\right)^2 - x \tilde h'_0(x) + \tilde h_0(x) \right]=0 \;.
\end{eqnarray}
Using the Painlev\'e II equation (\ref{PII}), one can indeed check that $\tilde h_0(x) = \int_x^\infty q^2(u) {\rm d}u$ is solution of this equation (\ref{PII_bis}). Note that to check this, it is useful to use the identity 
\begin{eqnarray}\label{identity}
\tilde h_0(x) = \int_x^\infty q^2(u) {\rm d} u = (q'(x))^2 - (q(x))^4 - x (q(x))^2 \;.
\end{eqnarray}

What is the first correction to the limiting form in Eq. (\ref{tilde_f0}) ? Unfortunately, in the soft edge limit, it turns out the Schlesinger equations  (\ref{recurrence_dotS}) do not allow to determine easily this first correction -- while they were very helpful in the hard edge scaling limit. An alternative way to compute this first correction is to analyze directly the Painlev\'e V equation in (\ref{Painleve5}). By inspection of this equation (\ref{Painleve5}), we conjecture that 
the first correction to Eq. (\ref{tilde_f0}) is of the form
\begin{eqnarray}\label{HN_kext}
H_N(t) = -(m x_-) \left(\tilde h_0(x) N^{2/3} + \tilde h_1(x) N^{1/3}\right) + {o}(N^{1/3}) \;.
\end{eqnarray}
By inserting this expansion (\ref{HN_kext}) in Eq. (\ref{Painleve5}) we obtain that $\tilde h_1$ satisfies the following linear differential equation
\begin{eqnarray}\label{eq_h1}
2 \tilde h_1 \tilde h'_0 +2 (\tilde h_0 + h'_0 (3 \tilde h'_0 -2 x)) \tilde h'_1 + \tilde h''_0 \tilde h''_1= 0 \;,
\end{eqnarray}
where $\tilde h_0(x)$ is given in Eq. (\ref{identity}). Of course, we know that $\tilde h_1(x) \to 0$ when $x \to +\infty$ (\ref{HN_kext}). In addition, from Eq. (\ref{eq_h1}) one can show, using Eq.~(\ref{identity}) and the large argument behavior in (\ref{PII}) $q(x) \sim {\rm Ai}(x) \sim x^{-1/4} e^{-(2/3)x^{3/2}}/(2 \sqrt{\pi})$ 
that $\tilde h_1(x) \sim A \, x^{-1/2} e^{-(4/3) x^{3/2}}$. But, of course, the equation (\ref{eq_h1}) being linear, the amplitude $A$ can not be determined from this analysis. One way to determine it would be to analyze the OP system (\ref{eq:def_OP}) in the limit when $t$ is far from the left edge, i.e., for $(N x_- -t)\gg N^{1/3}$ (corresponding to the left large deviation tail of $\lambda_{\min}$~\cite{Katzav10}), and then match this result with the typical regime, for $|N x_-- t| \sim {\cal O}(N^{1/3})$. This program was carried out in detail in a similar albeit different context, involving discrete OPs in Ref. \cite{Schehr13}. Since we are interested in the first correction, this actually requires a very precise (and tedious) analysis of this regime $(N x_- -t)\gg N^{1/3}$ which goes beyond the scope of the present paper. Hence our result for the first correction $\tilde h_1(x)$ in Eq. (\ref{eq_h1}) does determine this function only up to a constant. We have not found any simple solution to this equation (\ref{eq_h1}), which could indicate that the corrections to scaling in this case are actually more complicated, as found recently in the case of real Wishart matrices \cite{Ma12}.

\section{Conclusion}\label{section:Conclusion}

To conclude, we have provided a direct computation of the cumulative distribution $F_N(t)$ of the smallest eigenvalue of complex Wishart random matrices (\ref{Pjoint}), for arbitrary parameter $a \geq 0$. This was done by studying a set of semi-classical orthogonal polynomials as defined in Eq. (\ref{eq:def_OP}) for which we derived (i) the Schlesinger (\ref{schlesinger_intro}) and (ii) the Laguerre-Freud (\ref{LF_intro}) equations. By combining these equations, we showed that $F_N(t)$ can be expressed in terms of a solution of a Painlev\'e V equation (\ref{Painleve5}), thus recovering the result of Tracy and Widom \cite{Tracy94b} using a quite different method. In the large $N$ limit, $F_N(t)$, properly shifted and scaled, converges to a limiting distribution $F_{\infty}(x)$ which can be expressed in terms of a solution of a Painlev\'e III equation (\ref{PainleveIII}) in the hard edge limit (corresponding to $a ={\cal O}(1)$) and of a Painlev\'e II equation (\ref{PII_bis}) in the soft edge limit (corresponding to $a = {\cal O}(N)$). Furthermore, we have computed explicitly the first correction to the limiting distribution when $N \to \infty$. In the hard edge case (\ref{conjecture_finite_N}), we confirmed a conjecture by Edelman, Guionnet et P\'ech\'e in Ref. \cite{Edelman14}. In this case, the first correction can be simply understood as a correction to the scale of the fluctuations of $\lambda_{\min}$, see Eq. (\ref{eq:correction}). On the other hand, in the soft edge limit, we found that this correction is a solution of a second order linear differential equation with varying coefficients (\ref{eq_h1}). Solving this equation remains a challenging open problem, which could suggest that the first correction does not correspond to a simple shift or rescaling of the scaling variable, as found for the hard edge (\ref{eq:correction}). This could be reminiscent of the result found for real Wishart matrices \cite{Ma12} and certainly deserves further investigations.

\section*{Acknowledgments}

We would like to thank Y. Chen, P. J. Forrester and N. S. Witte for useful correspondence and discussions.

\appendix

\section{Some useful integrals involving the orthogonal polynomials}\label{appendix:Turan}

To compute the terms which enter the Laguerre-Freud equations, we need to compute several integrals containing the OPs $\pi_k(x)$ (\ref{eq:def_OP}). 
To perform these computations, we will follow the method developed by Belmehdi and Ronveaux \cite{Belmehdi94}, using Tur{\'a}n determinants. We introduce four types of integrals 
\begin{eqnarray}
\left\{
\begin{array}{ccl}
I_{m,k} &=& \braket{\lambda^m}{\pi_{k}^2},\\
\\
J_{m,k} &=& \braket{\lambda^m}{\pi_{k}\pi'_{k}},\\
\\
K_{m,k} &=& \braket{\lambda^m}{\pi_{k}\pi_{k+1}},\\
\\
L_{m,k} &=& \braket{\lambda^m}{(\pi_{k+1}\pi_k)'} \;,
\end{array}
\right.
\end{eqnarray}
for $m=0,1$ and $2$, while $k$ is an arbitrary integer. Using the recurrence relation (\ref{3term_rec}), we can calculate the integrals $I_{m,k}$'s and the $K_{m,k}$'s:
\begin{eqnarray}
\left\{
\begin{array}{ccl}
I_{0,k} &=& \braket{1}{\pi_{k}^2}=h_k,\\
\\
I_{1,k} &=& \braket{\lambda}{\pi_{k}^2}=S_k h_k,\\
\\
I_{2,k} &=& \braket{\lambda^2}{\pi_{k}^2}=(R_{k+1}+S_k^2+R_k)h_k,\\
\\
K_{0,k} &=& \braket{1}{\pi_{k}\pi_{k+1}}=0,\\
\\
K_{1,k} &=& \braket{\lambda}{\pi_{k}\pi_{k+1}}=h_{k+1},\\
\\
K_{2,n} &=& \braket{\lambda^2}{\pi_{k}\pi_{k+1}}=(S_{k+1}+S_k)h_{k+1}.
\end{array}
\right.
\end{eqnarray}
The terms $J_{0,k}$, $J_{1,k}$ and $L_{0,k}$ can easily be calculated using that $\pi_k$'s are monic OPs:
\begin{eqnarray}
\left\{
\begin{array}{ccl}\label{J0J1L0}
J_{0,k} &=& \braket{1}{\pi_{k}\pi'_{k}}=0, \\
\\
J_{1,k} &=& \braket{\lambda}{\pi_{k}\pi'_{k}}=k h_{k},\\
\\
L_{0,k} &=& \braket{1}{(\pi_{k+1}\pi_k)'}=(k+1)h_k.
\end{array}
\right.
\end{eqnarray}
The three last quantities, $J_{2,k},L_{1,k}$ and $L_{2,k}$, require more work. For this purpose, we introduce a new object: the Tur\'an determinant $\mathscr{T}$ defined as:
\begin{eqnarray}
\mathscr{T}_{k+1}=\pi_{k+2}\pi_{k}-\pi_{k+1}^2.
\end{eqnarray}
Using the three term recurrence (\ref{3term_rec}), we find a recurrence for the Tur\'an determinant
\begin{eqnarray}
\!\!\mathscr{T}_{k+1}&=&(\lambda \pi_{k+1}-S_{k+1}\pi_{k+1}-R_{k+1}\pi_{k})\pi_{k}-\pi_{k+1}(\lambda \pi_{k}-S_{k}\pi_{k}-R_{k}\pi_{k-1}),\\
&=& R_k \pi_{k+1}\pi_{k-1}+(S_k-S_{k+1})\pi_{k+1}\pi_k-R_{k+1}\pi_k^2,\\
&=& R_k\mathscr{T}_{k}+F_k,
\end{eqnarray}
where we have introduced the auxiliary quantity $F_k=(S_k-S_{k+1})\pi_{k+1}\pi_k+(R_k-R_{k+1})\pi_k^2$.

Finally, we can write, using $R_k=h_k/h_{k-1}$ and the initial condition $\mathscr{T}_{1}=F_0$ 
\begin{eqnarray}
\mathscr{T}_{k+1}=h_k \sum_{i=0}^{k} \frac{F_i}{h_i},
\end{eqnarray}
This equation allows us to rewrite
\begin{eqnarray}\label{scalTuran}
\braket{\lambda^m}{\mathscr{T}_{k+1}'}=h_k\sum_{i=0}^{k}h_i^{-1}\left((S_i-S_{i+1})L_{m,i}+2 (R_i-R_{i+1}) J_{m,i}\right),
\end{eqnarray}
where the prime denotes a derivative w.r.t. $x$. On the other hand, by using the definition of the Tur\'an determinant we find
\begin{eqnarray} \label{scaln2'}
\braket{\lambda^m}{\mathscr{T}_{k+1}'}=\braket{\lambda^m}{(\pi_k \pi_{k+2})'}-2 J_{m,k+1}.
\end{eqnarray}
We obtain finally
\begin{equation} \label{scal_pipi2'}
\braket{\lambda^m}{(\pi_k \pi_{k+2})'}=2 J_{m,k+1}+h_k\sum_{i=0}^{k}h_i^{-1}\left((S_i-S_{i+1})L_{m,i}+(R_i-R_{i+1})2 J_{m,i}\right).
\end{equation}
We now have all the ingredients needed to calculate the last three integrals.

Using the three term recurrence and the orthogonality condition, we can express $L_{1,k}$ as
\begin{eqnarray}
L_{1,k}=\braket{\lambda}{\pi'_{k+1}\pi_k}&=&S_k\braket{\pi_k}{\pi'_{k+1}}+R_k\braket{\pi_{k-1}}{\pi'_{k+1}}, \\
&=& S_k (k+1) h_k +R_k \braket{\pi_{k-1}}{\pi'_{k+1}}.
\end{eqnarray}
From Eq. (\ref{scal_pipi2'}) for $m=0$ , we can express the last scalar product in terms of integrals that we have computed before. Using Eq.~(\ref{J0J1L0}), we find
\begin{eqnarray}\label{L1}
L_{1,k}&=& S_k (k+1) h_k +R_k h_{k-1} \sum_{i=0}^{k-1}h_i^{-1} (S_i-S_{i+1})(i+1) h_i = h_k \sum_{i=0}^k S_i.
\end{eqnarray}
Using the three term recurrence and the orthogonality condition, we can express $J_{2,k}$ as
\begin{eqnarray}
J_{2,k}&=&\braket{\lambda^2}{\pi_{k}\pi'_{k}} = S_k \braket{\lambda}{\pi_{k}\pi'_{k}} + R_k \braket{\lambda}{(\pi_{k-1}\pi_{k})'} \nonumber \\
&=& S_k J_{1,k}+ R_k L_{1,k-1}.
\end{eqnarray}
Finally, we have from Eqs. (\ref{R_en_h}), (\ref{J0J1L0}) and (\ref{L1}),  
\begin{eqnarray}
J_{2,k}=h_k ( k S_k+\sum_{i=0}^{k-1} S_i).
\end{eqnarray}
Similarly, using again the three term recurrence and the orthogonality of the family, we can express $L_{2,k}$ as
\begin{eqnarray}
L_{2,k}&=&\braket{\lambda^2}{(\pi_{k+1}\pi_k)'} \nonumber \\
&=& \braket{\lambda^2}{\pi_{k+1}\pi_k'} + \braket{\lambda}{\pi_{k+1}\pi_{k+1}'} +S_k \braket{\lambda}{\pi_{k}\pi_{k+1}'}+R_k \braket{\lambda}{\pi_{k-1}\pi_{k+1}'} \nonumber \\
&=& k h_{k+1}+ J_{1,k+1}+S_k L_{1,k}+R_k \braket{\lambda}{(\pi_{k-1}\pi_{k+1})'}.
\end{eqnarray}
Using Eq. (\ref{scal_pipi2'}) for $m=1$, we can express the last scalar product in terms of integrals that we have computed before.
\begin{eqnarray}
L_{2,k}&=&(2k+1)h_{k+1}+S_k h_k \sum_{i=0}^k S_i + R_k 2k h_k \nonumber \\
&+&h_k \sum_{i=0}^{k-1}\left[ (S_i-S_{i+1})\sum_{j=0}^i S_j+2i(R_i-R_{i+1})\right].
\end{eqnarray}
We can simplify the telescoping sums and find
\begin{eqnarray}
L_{2,k}=(2k+1)h_{k+1}+h_k\left(2\sum_{j=1}^k R_j+\sum_{j=0}^k S_j^2\right).
\end{eqnarray}
Equipped with the computations of these integrals, we can now derive the Laguerre-Freud equations.

\section{Details related to the Laguerre-Freud equations}\label{appendix:LF}

The derivation presented here follows the one of Ref. \cite{Forrester07} where slightly different OPs were considered. To derive the Laguerre-Freud equations, we start from equation (\ref{LF01}) :
\begin{eqnarray}
\left\{
\begin{array}{rcl}\label{LF01_A2}
&&R_{k+1}+R_k+S_k(S_k-a-t-2k-2)+t(2k+1+a) = 2\displaystyle{\sum_{j=1}^{k-1}} S_j, \\
\\
&&(S_{k+1}+S_k-3-a-t-2k)R_{k+1} = \displaystyle{2\sum_{j=1}^k R_j+\sum_{j=0}^k S_j^2 - t \sum_{j=0}^k S_j.}
\end{array}
\right.
\end{eqnarray}
It is useful to introduce the following variables
\begin{eqnarray}\label{def_omega}
\left\{
\begin{array}{ccl}\label{def_omega_A2}
\theta_k&=&2k+1+a-S_k,\\
\\
\omega_k&=&-R_k-\zeta_k. 
\end{array}
\right.
\end{eqnarray}
In terms of these variables, we can rewrite the first equation of (\ref{LF01_A2}) as
\begin{eqnarray}\label{LF_omega1}
\omega_{k+1}+\omega_k&=&(t-S_k)\theta_k,
\end{eqnarray}
and the second equation of (\ref{LF01_A2}) as
\begin{eqnarray}\label{LF_omega2}
(t-S_k)(\omega_k-\omega_{k+1})=\theta_{k-1}R_k-\theta_{k+1}R_{k+1}.
\end{eqnarray}
Similarly, Eq.~(\ref{LF_omega2}) can be written as
\begin{eqnarray}
(\lambda+t-S_k)(\lambda+\omega_{k+1}-\omega_{k})&=&\theta_{k+1}R_{k+1}-\theta_{k-1}R_k+\lambda(\lambda+t-S_k+\omega_{k+1}-\omega_{k})\nonumber \\
&=&\theta_{k+1}R_{k+1}-\theta_{k-1}R_k+\lambda(\lambda+t+R_k-R_{k+1}) \nonumber \\
&=&(\theta_{k+1}-\lambda)R_{k+1}-(\theta_{k-1}-\lambda)R_k+\lambda(\lambda+t)\;.\label{LFO1_omega}
\end{eqnarray}
We also rewrite Eq.~(\ref{LF_omega1}) as
\begin{eqnarray}\label{LFO2_omega}
(-\lambda^2+\lambda(2k+a-t)+1+\omega_{k+1}+\omega_k)&=&(\lambda+t-S_k)(\theta_k-\lambda)\,.
\end{eqnarray}
Finally, by multiplying together Eqs. (\ref{LFO1_omega}) and (\ref{LFO2_omega}), we have, with the notation
 $\Omega_k(\lambda)=-\frac{\lambda^2}{2}+\frac{\lambda}{2}(2k+a-t)+\omega_k$, 
\begin{eqnarray}
(\Omega_{k+1}(\lambda)-\Omega_{k}(\lambda))(\Omega_{k+1}(\lambda)+\Omega_k(\lambda))&=&(\theta_k-\lambda)((\theta_{k+1}-\lambda)R_{k+1} \nonumber \\
&-&(\theta_{k-1}-\lambda)R_k+\lambda(\lambda+t)).
\end{eqnarray}
For $\lambda=0$, we find
\begin{eqnarray}
\omega_{k+1}^2-\omega_k^2&=&R_{k+1}\theta_{k+1}\theta_k-R_k \theta_k \theta_{k-1}.
\end{eqnarray}
Using the initial condition $R_0=\omega_0=0$, we can solve this equation and find
\begin{eqnarray}\label{theta_km}
\omega_k^2&=&R_k \theta_k \theta_{k-1} \;.
\end{eqnarray}
For $\lambda=-t$, we find
\begin{equation}
\Omega_{k+1}(-t)^2-\Omega_{k}(-t)^2=(\theta_k+t)(\theta_{k+1}+t)R_{k+1}-(\theta_{k-1}+t)(\theta_k+t)R_k.
\end{equation}
Again, using the initial condition $R_0=\omega_0=0$, we can solve this equation and find
\begin{eqnarray}
\Omega_{k}(-t)^2&=&(\theta_{k-1}+t)(\theta_k+t)R_k+\left(\frac{t}{2}(2k+a)\right)^2.
\end{eqnarray}
Finally, using the explicit expression of $\Omega_{k}(-t)$, this equation can also be written as
\begin{eqnarray}\label{theta_km2}
(\omega_k-kt)( \omega_k-(k+a)t)&=&R_k(\theta_k+t)(\theta_{k-1}+t)\;.
\end{eqnarray}
Thanks to Eq.~(\ref{theta_km}), we can substitute $\theta_{k-1}$ in (\ref{theta_km2}). Using the definition of $\omega_k$ in (\ref{def_omega}), Eq. (\ref{theta_km2}) yields a quadratic equation for $\omega_k$ in terms of $\theta_k$ and $\zeta_k$
\begin{eqnarray}
\omega_k^2 -\theta_k (\theta_k-a-2k+t) \omega_k -\theta_k \left(k t(k+a)+(\theta_k+t)\zeta_k\right) = 0 \;,
\end{eqnarray}
which yields the equation given in the text in Eq.~(\ref{LF02_bonne_limite}).

\section{Numerical simulations}\label{appendix:Numeric}

\begin{figure}[ht!]
\begin{center}
\resizebox{120mm}{!}{\rotatebox{-90}{\includegraphics{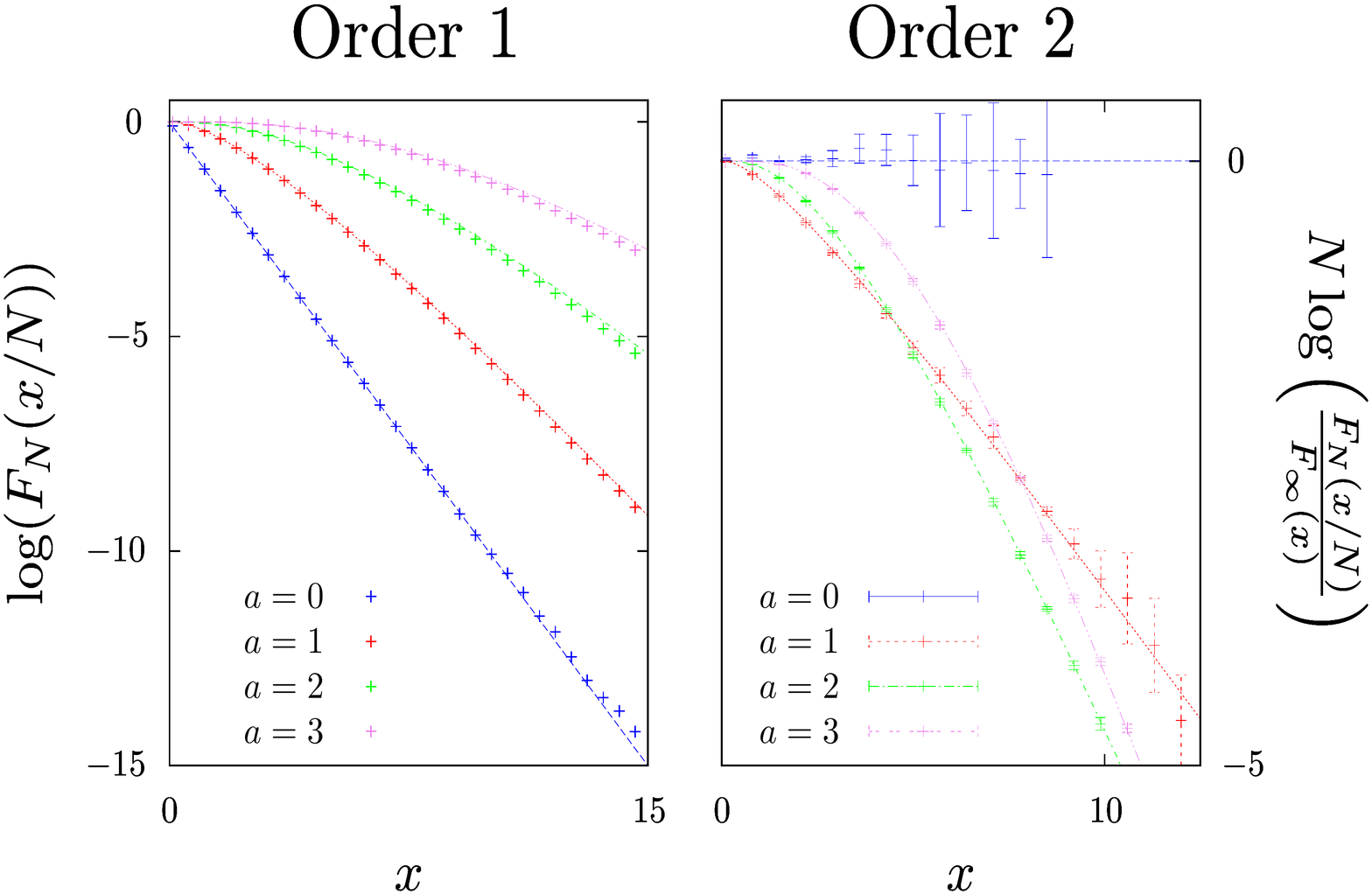}}}
\caption{Comparison between our formula (\ref{Dvpt_F}) using (\ref{fdetBessel}) (dashed curves) and data obtained by sampling $3.10^7$ independent Wishart matrices of size $N=50$ (dots) for different values of $a=0,1,2$ and $3$. The left panels presents a comparison between $\log(F_N(x/N))$ and $\log (F_\infty(x))$ given in (\ref{Finfty(x)}). The right panel illustrates the finite $N$ correction by comparing $N\log\left(\frac{F_N(x/N)}{F_\infty(x)}\right)$ to $\frac{a}{2} f(x)$, see Eq. (\ref{correction_app}).}
\label{fig:Fn_ordre_2}
\end{center}
\end{figure}

In this section, we present a numerical check of our formula for the first finite $N$ corrections to the limiting density of $\lambda_{\min}$ in the hard edge case, i.e., keeping $a$ finite while $N \to \infty$. Our simulations have been carried out for integer values of the parameter $a$ for which the function $f(x)$ in Eq. (\ref{Finfty(x)}) can be computed explicitly in terms of Bessel $I$ functions~\cite{Forrester94}:  
\begin{eqnarray}\label{fdetBessel}
f(x) = -x \frac{\det{[I_{j-k+2}(2\sqrt{x})]_{1\le j,k \le a}}}{\det{[I_{j-k}(2\sqrt{x})]_{1\le j,k \le a}}}\,, \; a \; {\in} \; {\mathbb N}^* \;,
\end{eqnarray}
and $f(x)=-x$ for $a=0$. 
In Fig. (\ref{fig:Fn_ordre_2}) we show a plot of $\log F_N(x/N)$ obtained by sampling $3.10^7$ independent Wishart matrices of size $N=50$ (dots) for different values of $a=0,1,2$ and $3$. The comparison with $\log F_{\infty}(x)$, evaluated exactly from Eqs. (\ref{Finfty(x)}) and (\ref{fdetBessel}) shows a good agreement between theory and numerical simulations. The main objective of these simulations is to test the formula for the first order correction, given in Eq. (\ref{Finfty2}) which can also be written as
\begin{eqnarray}\label{correction_app}
\log{\left(\frac{F_N(x/N)}{F_{\infty}(x)} \right)} = \frac{1}{N} \frac{a}{2} f(x) + o(1/N)\;.
\end{eqnarray}
In the left panel of Fig. \ref{fig:Fn_ordre_2}, we show a plot of $N \log{\left[{F_N(x/N)}/{F_{\infty}(x)} \right]}$ as a function of $x$ and compare it to $\frac{a}{2} f(x)$, where $f(x)$ is given in Eq. (\ref{fdetBessel}). This comparison illustrates the accuracy of the asymptotic formula characterizing the finite $N$ correction to $F_{\infty}(x)$.

%\newpage

\end{document}